*Anna Denkowska*
E-mail: anna.denkowska@uek.krakow.pl
ORCID ID: https://orcid.org/0000-0003-4308-8180

Department of Mathematics, Cracow University of Economics, Rakowicka 27, 31-510 Kraków,

*Stanisław Wanat*
E-mail: wanats@uek.krakow.pl
ORCID ID: https://orcid.org/0000-0001-6225-4207

Department of Mathematics, Cracow University of Economics, Rakowicka 27, 31-510 Kraków,

# Development and similarity of insurance markets of European Union countries after the enlargement in 2004

**Abstract**

**Research background:** The enlargement of the European Union to new countries in 2004 launched mechanisms supporting the development of various social and economic areas, as well as levelling the differences between the Community members in these areas. This article focuses on the insurance sector.
**Purpose of the article:** The main purpose is to analyze the development and similarity of the insurance markets of old and new members of the European Union after the enlargement in 2004. In this work, multi-dimensional analysis is used to reveal similarities and differences between the individual countries in these groups of countries, in particular in 2005, 2009 and 2015, 2018.
**Methods:** We study simultaneously the development of the „Old" and „New" EU countries, using Hellwig's development measure, taking into account a set of certain characteristics: population of a given country, GDP, number of insurance companies, total assets, gross written premium on Life and P&C and Health, penetration index divided into Life and P&C and Health, insurance density index separated into Life and P&C and Health, concentration indicator broken by Life and P&C. We analyze the similarity of countries using three statistical methods of unsupervised classification: the Ward's method, the k-means method and the partitioning among medoids (PAM).

**Findings & Value added:** Between the groups of the old and the new European Union after 2004 there was a fairly large variation in the characteristics of insurance. In general, in each group, insurance markets develop in a different way. Markets did not align after 2004.

**Introduction**

In 2004 The European Union consisting of 15 countries such as Austria, Belgium, Denmark, Finland, France, Greece, Spain, the Netherlands, Ireland, Luxembourg, Germany, Portugal, Sweden, Great Britain and Italy also referred to as the "Old Union" (Old EU) has been expanded to include countries Cyprus, the Czech Republic, Estonia, Lithuania, Latvia, Malta, Poland, Slovakia, Slovenia and Hungary (New EU). Then Bulgaria, Romania and Croatia joined the EU. UK is no longer a member of the Union from February 1st 2020. We want to answer questions related to the development in the insurance sector of the "Old Union" countries in comparison to the countries that joined after 2004. Did the EU enlargement affect any of these groups of countries? To what extent were the insurance markets of the different countries similar? Or are the similarities due to the structural changes that followed the enlargement? Has the community goal been achieved yet? Have the differences between the member countries been resolved, equalized?
The analytical part consists of two steps. The first is development analysis with Hellwig's method ranking, the second is to determine groups of similar insurance markets in European Union countries.
The development analysis is done for the entire sector and divided into life and non-life insurance. In this analysis, we use, among others, the following measures: the number of insurance companies, the market concentration indicator, gross premiums written, total assets, the insurance density ratio, the insurance penetration ratio, the life and non-life insurance share of the total insurance market. We use multivariate statistical analysis methods i.e. principal components analysis, statistical methods of unsupervised classification. The development of the insurance market depends not only on factors resulting from the principles and manner of its functioning. Economic and social conditions also have an impact on its structure. In each country, features such as the level of economic development of the country, the wealth of households, or the amount of insurance premiums collected are quite different. In addition, public awareness and insurance tradition is very important. By comparing all these factors for European countries, you can obtain a very large set of information on the overall insurance structure in the EU.

The calculation is based on statistical data published by the European Insurance and Occupational Pensions Authority (EIOPA), the Schweizerische Rückversicherungs-Gesellschaft (Swiss Re) and the European Statistical Office (Eurostat). All calculations were made in the R program.

Finally, the conclusions of the analysis are presented. Tracking development together with similarities and differences we pay attention to specific periods on the timeline. 2005 is the first full year of membership of Poland and other 9 countries in the European Union. Due to a lack of available data Lithuania had to be left out from the analysis. The year 2009 is a period of "struggle" against the effects of the global economic crisis that began in mid-2007. 2015 is the year when an immigrant crisis starts in Europe. In turn, 2018 provides the most current data for all variables used in the analysis.

The novelty of the present article consists in the study the development and similarity of the insurance sectors of EU countries using Hellwig's development measure and cluster analysis methods.

**Literature review**

The fifth stage of EU enlargement was the largest in terms of the number of countries joining the Union and hence in terms of population. The inclusion of 10 new countries into the EU changed the dynamics of the EU's development, created new opportunities for economic, economic, political, cultural and legal development Dimiter & Toshkov (2017). The literature review should begin by pointing to a certain gap in research on the insurance sector in the context of the enlargement of the European Union. Despite the fact that in the literature we find numerous reports and articles in which the authors examine the effects of the enlargement, it is difficult to resist the impression that the insurance sector is treated too superficially. Every year, the European Commission produces reports in which the overall economic, financial and political situation is analyzed. In the EU report (2006) we can find a summary of the first two years of integration. In the ECD (2009) we find an analysis of the economic achievements of the Union and the challenges five years after integration. Lepesant (2014) is a review of a decade of enlargements 2004-2014. The ECDG FISMA report (2019) assesses economic activity, describes changes in financial sector policy and focuses on the cross-border dimension of macro-prudential policy in the EU banking sector where the European Systemic Risk Board (ESRB) plays an important role, in 2018 and in the first quarter of 2019. The report analyzes the growing use of artificial intelligence (AI) applications in financial services. The development of the insurance sector is briefly summarized based on EIOPA (2018). The ECDG FISMA report (2020) states, inter alia, that "The insurance sector has performed well, but the mid-to-long-term market outlook looks more challenging." It briefly summarizes how the Insurance Sector coped with the first stage of the Pandemic that started in 2019. In MAPFRE Economics (2020), a ranking of the largest European insurance groups by premium volume and of growth dynamics over the past decade was established. Among the articles on the insurance sector, we can find a work in which Ertl, M. (2017) examines the relationship between economic growth and insurance activity, analyzing the development of life and property insurance in 1994-2014. On the other hand, Wanat et al. (209) analyze causal relations between the insurance market development and economic growth in ten transition European Union member countries in the period between 1993 and 2013. The analysis is conducted with the use of the bootstrap panel causality approach. Various types of dependencies between economic growth and the insurance market development (both in terms of the global insurance market and in the division into life insurance and non-life insurance) are identified in the study. Kwon and Wolfrom (2016) discuss analytical tools used by macroprudential supervision and insurance regulators. Kozarevic et al. (2013) analyze the development of the insurance sector in the Western Balkan Countries and the process of integration with the European Union of countries belonging to it and applying for accession. Nevertheless, Cummins and Rubio-Misas (2018) highlight gaps in the literature in the area of research into the insurance sector. At the same time, they analyze the impact of integration on the effectiveness of the life insurance markets in the European Union (EU) in the post-deregulation period 1998-2011. Alvarez and Makunin (2018) przedstawiają raport dotyczący branży ubezpieczeniowej w wybranych krajach regionu CEE. EIOPA (2015, 2016, 2017, 2018, 2019) has statistical data. There is no detailed analysis of the development and similarity of European Union insurance markets and no analysis of changes taking place in the insurance sector in the fifteen-year period since 2004. Our study therefore supplements the literature. The work uses taxonomic methods to investigate the development and similarity of the insurance markets of the European Union countries.

**Research methodology**

The enlargement of the European Union with new countries in 2004 was supposed to launch mechanisms supporting the development of various socio-economic areas, as well as eliminating the differences between the members of the Community in these areas. The paper examines whether such consequences of enlargement can be observed in the insurance markets of EU countries. We hypothesize that in general, in each group of old and new EU members, insurance markets develop in a different way. Markets did not align after 2004. Between the groups of the old and the new European Union after 2004 there was a fairly large variation in the characteristics of insurance. When examining the state of development of European Union insurance markets, it can be seen that still a long way separates Central and Eastern European countries from reaching western market levels. Nevertheless, the insurance industry is developing very dynamically in these countries.

We conduct the analysis using selected methods of multivariate statistical analysis. We study the development of the insurance sectors of EU countries after 2004, constructing the so-called development path: $g_i^{2004}, g_i^{2005}, \ldots, g_i^{2018}$, where $g_i^t, i = 1, \ldots 27, t = 2004, \ldots 2018$ is a measure of Hellwig's development measure (Hellwig (1968)). This measure is made up of appropriately scaled Euclid's distances between the multidimensional vector of normalized values of variables characterizing the development of the insurance sector in a given country $i$ in a given year $t$ from the so-called development pattern. Due to the dynamic nature of the analysis, we take a multidimensional vector as the development pattern $z_{dp} = [z_{max,1}, z_{max,2}, \ldots, z_{max,k}]$, the coordinates of which are the maximum values of the normalized variables (after prior conversion of the destimulant into stimulants) taken in all countries $i = 1, \ldots 27$ and t = 2004, ... 2018, i.e. $z_{max,j} = \max_t \max_i z_{i,j}^t$ (Zeliaś (2000)).

On the other hand, we examine the similarity of insurance markets with the use of statistical methods of unsupervised classification. Clustering is conducted by means of hierarchical methods in which groups are created recursively by linking together the most similar objects (Ward's method is applied here). Other methods of division, i.e. the k-means method and the partitioning among medoids (PAM) method proposed by Kaufman and Rousseeuw (1990) are also used. In both cases, after making the initial decision about the desired number of groups, objects are allocated in such a way that the relevant criterion is met. For the k-means method the allocation of objects should minimize a within-group variance. In the PAM method the representatives of groups (medoids) are selected at each step of the analysis, and then the remaining objects are allocated to the group which includes the closest medoid. The former method is more robust to outliers than the k-means method, because it minimizes the sum of dissimilarities instead of the sum of squared Euclidean distance. In order to evaluate the optimal number of clusters in the data, we use internal validity indexes: Calinski Harabasz pseudo F statistics (Calinski and Harabasz, 1974), the average silhouette width (Kaufman and Rousseeuw, 1990), the Dunn index (Dunn, 1974), and Xie and Beni's (1991) index. The final classification of objects is, therefore, the result of the comparison of the results of respective grouping algorithms.

We analyze both the development and the similarity of the insurance sectors on the basis of the following diagnostic variables:
- Life insurance penetration[1] - indicates the level of development of life insurance sector in a country.
- Property & Casualty insurance penetration - indicates the level of development of property and casualty insurance sector in a country.
- Life insurance density[2] - indicates the level of development of life insurance sector in a country.
- Property & Casualty density - indicates the level of development of property and casualty insurance sector in a country.
- Average value of total insurance assets (i.e. total life, P&C and Health sector) per one insurance company[3] - indicates the "strength" of insurers from a given country.
- Investment to GDP ratio - indicates the 'strength' of the country's insurance sector.
- Market share of the top 5 life insurance groups - indicates the competitiveness of the life sector in a given country.

---

[1] Penetration rate is measured as the ratio of premium underwritten in a particular year to the GDP.
[2] Insurance density is calculated as the ratio of total insurance premiums to whole population of a given country.
[3] Total insurance assets (i.e. total life, P&C and Health sector) of a given country divided by the number of insurers in that country.

- Market share of the top 5 property and casualty insurance groups - indicates the competitiveness of the P&C sector in a given country.
- Number of companies on total market per 1 million inhabitants - indicates the competitiveness of the insurance sector in a given country.

The values of these variables for individual countries and years were obtained from databases Insurance Europe[4] (https://www.insuranceeurope.eu/about-us). To the best of the authors' knowledge, this type of approach to study the development and similarity of the insurance sectors of EU countries, using development paths and cluster analysis methods, has not been used in the literature. Its advantage is the inclusion in the analyzes of many aspects of the insurance sectors represented by individual diagnostic variables. A certain limitation is the selected set of diagnostic variables, which was partially determined by the availability of complete statistical data.

**Results**

*1. Market development analysis*

In the first step of our research, we analyze the development of EU insurance markets in the years 2004 - 2018. We study simultaneously the development of the "old" and "new" EU countries, taking into account a set of certain characteristics: population of a given country, number of insurance companies, total assets, GDP, gross written premium on Life and P&C and Health, penetration index divided into Life and P&C and Health, insurance density index separated into Life and P&C and Health and concentration indicator broken by Life and P&C. In the analysis, we pay particular attention to the years 2005, 2009, 2015 and 2018. In most cases, we present two graphs, for old and new EU countries in one coordinate system for trends comparison. On the right is the unit for countries that have been in the Union since 2004, on the left for countries of the old Union. We start the analysis of the differences between the old and the new EU countries by comparing the *number of population* in Fig. 1.1 In the countries of the old Union we observe an increase in population caused both by immigration from outside Europe as well as from the countries of the new Union. On the other hand, in the New Union countries the population evolution does not show a single trend. We can see that since 2004-2007 the population has been falling, then it fluctuated in the range of 70600-70650, and then it falls below 70500, which has an impact on the overall level of development of the insurance market according to the principle: the larger the population, the greater the potential.

In what follows, we analyze the *number of insurance companies* in the old and new EU. The number of insurance companies in each group decreases year by year, as shown in Figure 1.2.The number of insurance companies per million inhabitants is one of the most basic indicators that can be used to determine the development level of the insurance sector and the competition of insurance markets between different countries. The number of insurers involved in life-insurance show the level of development of this sector. Among the Eastern and Central European countries the Polish insurance market is the largest one. The biggest share of the UE market is held by a group of insurance companies from the UK, France, Germany and Italy. The number of insurance companies of the Old UE shows a decreasing tendency while in the New UE it seems to stay at a stable level.

Another indicator is *total assets*. Insurance companies in the world determine their market share by the size the managed assets. The company's value is proved by its assets. The value of assets, i.e. of those managed by the institution, results directly from investment activities. First of all, it depends on the profits made by pension and investment funds managed by insurance companies. For the individual customer, the value of assets is important because it affects e.g. the amount of their future retirement pension. For the economic market of a given country itself, the assets of the insurers are a powerful capital that supplies this market. Experts estimate that in the next five years the role of insurance savings in the deposits of the population will increase. This means that the assets accumulated in the insurance services segment will start to play a dominant role in determining the condition of financial markets. As we can see from the data analysis, the forecasts are coming true. After the reaction in the 2007-2009 periods in the old Union and in 2007 and in 2010, the value of assets increases (Fig. 1.3). The value of assets increases throughout the entire analyzed period except for the subprime crisis period. In Fig. 1.4 is a graphic summary of total assets for the two compared groups of countries.

*GDP* in both groups tends to grow (Fig. 1.5). In 2008 we observe a decline in GDP in both groups. During the sovereign debt crisis in European countries, GDP has remained at an equal level, since 2014 we have seen continuous growth. Economic development measured by GDP per capita is strongly associated with the development of the insurance market, which is measured by gross written premium. A low share of gross written premium in life insurance in the total premium is characteristic of countries with low GDP per capita. Figure 1.6 shows three graphs describing the change in the level of *gross written premium in Life, P&C and Health* activities. The gross premium written for life is the most variable. Although it should be noted that the "old" Union is more responsive during the 2008 crises and in 2011. Then the premium tends to increase. On the other hand, in the countries that later joined the EU, since their accession this premium has been rising, in the years of crises we see some stagnation, and then it decreases. This can be linked to the decline in population in this part of Europe. For the rest, premiums other than Life have a growing trend. The gross insurance premium is the basic measure of the situation of insurance companies and the insurance market as a whole. It allows the assessment of the development of the "insurance industry" and the structure of the insurance portfolio. By analyzing the growth rate of non-Life insurance premium collected by insurers, we confirm the dependence of the degree of development of a given insurance market on the amount of GDP.

---
[4] Insurance Europe is the European (re)insurance federation

In Fig. 1.7 and Fig. 1.8 we present an analysis of the premium structure: In the old UE countries, life is dominant and health is developing. In the new ones, property is dominant and health is minimal. You can see some small progress. We wonder if this is a loss of confidence in long-term insurance? During the crisis, the old ones' Life GPW decreased in force and in the new ones Life GPW increased.

The insurance *penetration* rate expresses the relation between the insurance premium and the gross domestic product. This indicator belongs to the group quantifying the role that insurance plays in the entire national economy. A higher ratio of the annual insurance premium to the GDP shows that the protection of property, life and health of the population is widely used, in other words it illustrates a higher insurance awareness in a given country. We note that the percentage of penetration of new EU countries varies depending on the type of insurance. P&C penetration is the highest, comparable to P&C penetration of the old EU countries. We observe that the nine countries' Life penetration decreases in 2008 and stays at a reduced level. For old countries it is variable and clearly decreases in 2008 and then in 2011, and then in 2016. P&C penetration drops until 2008, then increases to reach a minimum again in 2015 in both groups of countries. Penetration for Health has been stable for the countries of the old Union since 2006, but almost two times lower than Life and P&C penetrations (Fig. 1.9).

The insurance *density* indicator is a measure of the development of the insurance sector, more precise than the level of gross written premium. It is defined as the value of gross written premium per capita. Based on the information defining how much financial resources a statistical resident spends on insurance, we can draw conclusions regarding the degree of development of insurance awareness of the citizens of a given country. The density level, similarly to previous indicators, is the highest for Life, then for P&C and the lowest for Health. In the countries of the new Union it decreases in 2008 and remains at this level. For the old EU countries, we observe a decrease during the excessive debt crisis in the Eurozone. For P&C it has fluctuations and for Health it has a growing trend (Fig. 1.10).

*Investment portfolio* on domestic market (Fig. 1.11) after crisis and decrease in 2008 in the old Union shows a growing trend, while in the new Union it keeps decreasing.

*Concentration*, i.e. the market share, of five insurers is the information about how much of the market is owned by these five largest insurers in a given country. Lowering the value of concentration means increasing competition on the insurance market. An interesting situation took place for the old Union in 2014, concentration for Life decreased while for P&C it increased. For the new UE members, concentration begins to decline in 2008 and only in 2012 starts to rise again. (Fig. 1.12)

The ranking of the development paths of the old and new EU countries, determined using the Hellwig's method, is presented in Fig. 1.13. We observe no trends in both groups. The old EU countries, whose development is at a higher level, are marked in blue. The countries of the new Union whose level of development is much lower are marked in red. It appears clearly that in the analyzed period the development paths of the two groups of countries are separated, which proves that the integration in 2004 did not significantly change the level of development of the insurance sectors in any of these group. Throughout the entire period one country significantly stands out: Greece, whose level of development is much lower than that of other countries from the group of the old EU. Figure 1.14 presents the ranking (X axis) and Hellwig's development measure (Y axis) for 2005, 2009, 2015 and 2018. The average level of development of the new EU countries is constantly at the same level, for the old EU countries the average level of development slightly decreases. Change in Hellwig's development measure in 2018 compared to 2004 (in %) is shown in Fig. 1.15.

*2. Analysis of the similarity of insurance markets.*

In the second step of the research, we analyze the similarity of EU insurance markets in the years 2004 – 2018 with the use of statistical methods of unsupervised classification.

Clustering is conducted with the use of three methods: Ward's method, the k-means method and the partitioning among medoids (PAM). We present the dendrograms for Ward's method in Fig.2.1. The assessment of the quality of clustering is presented in Table 2.1 and index validation clustering is calculated with the assumption that the number of groups in not smaller than 2 and not larger than 6. Cluster analysis results are shown in Fig. 2.2. The Silhouette plots for this partition is shown in Fig. 2.3.

Analyzing the results for 2005 we can see that Silhouette index indicates, depending on the adopted method, the division into two or three groups is optimal, while the Caliński-Harabasz index prefers to focus on two groups, regardless of the method used by us. The remaining index, i.e. the Dunn index, depending on the method used, gives the optimal division into four groups (Ward's method) or five groups (k-means method) and three (PAM method). The Xie-Beni index, depending on the method used, gives the optimal division into four groups (PAM method) or five groups (k-means method and Ward's method). Eventually, we decide to consider division into two groups based on the k-means method and we get two clusters:
- Cluster 1: AT BE CY DE DK ES FI FR IE IT LU MT NL PT SE UK
- Cluster 2: CZ EE GR HU LV PL SI SK *(NEW EU without CY and MT but with GR)*

Analyzing the results analogously for 2009 we opt for a clustering in six groups based on PAM method:
- Cluster 1: AT DE NL
- Cluster 2: BE IT UK
- Cluster 3: CZ CY EE GR HU LV MT PL SI SK *(NEW EU with GR)*
- Cluster 4: DK FI IE LU SE
- Cluster 5: ES PT
- Cluster 6: FR

Based on the results for 2015 we choose the partitioning into three groups based on k-means method:
- Cluster 1: AT BE DE DK FR IT NL UK
- Cluster 2: FI IE LU SE
- Cluster 3: CY CZ EE ES GR HU LV MT PL PT SI SK *(NEW EU with ES, GR, PT)*

On the ground the results for 2018 we opt for the clustering in three groups based on Ward's method:
- Cluster 1: SK GR HU CZ SI PL EE LV CY MT PT ES AT *(NEW EU with GR, ES, PT, AT )*
- Cluster 2: DK LU DE BE FI NL IE SE
- Cluster 3: FR IT UK

**Discussion**

Insurance markets of individual countries develop under the influence of various factors, which can be divided into economic, demographic, cultural, social and structural. They can affect the insurance market both positively and negatively. Economic factors are a very important element in the development of life and non-life insurance. Insurance demand certainly largely depends on the amount of disposable income per capita - the higher the income, the higher the level of demand and development of life insurance, assuming a relatively low level of inflation. Low-income entities have a higher risk tolerance, so their insurance demand is low. On the other hand, entities that have larger assets at risk of loss report greater demand for insurance products.

Demand for insurance services also depends on the price of this service, i.e. the insurance premium. Other structural factors are also important in the development of insurance, e.g. political decisions regarding the promotion of competition from foreign insurers or the direct provision of insurance by the public sector. Tax credit for the purchase of pension insurance as well as the level of development of social security systems have a positive impact on the demand on the insurance market. In countries with low GDP per capita, security systems are usually underdeveloped. In richer countries, non-economic factors tend to limit the growth of Section I insurance, because a higher the GDP per capita is accompanied by an increase in benefits from the social security system, which in turn has a negative impact on the development of life insurance.

The development analysis carried out in this paper indicates the persistent differences in the level of development between the analyzed groups of the old and new EU countries. And although you can observe a growing trend of some variables, e.g. Total assets, GDP, the average level of development remains at the same level, because e.g. the number of companies is decreasing. We also note that selected variables such as gross written premium, penetration or density react dynamically to crises. The analysis of the similarity also does not indicate any real integration of the insurance sector. For a number of years the new EU countries have not become similar to the old ones. In 2009, due to the subprime crisis, we also observe a breakdown of the group of old EU states. Comparing our work with what has been done in the existing literature, we observe that there was no such multidimensional and overall analysis of the insurance sector for the last 15 years in EU countries. Ususally, researchers concentrate either on few selected countries, or consider only statistical data. For instance, in Cummins and Rubio-Misas (2019) the authors examine the impact of integration on the efficiency of European Union life insurance markets for the post-deregulation period 1998-2011. Alvarez and Makunin (2018) present a report on the statistical analysis for the period 2013-2017 of the insurance industry in selected countries of the CEE region (Hungary, Czech Republic, Slovakia, Slovenia, Latvia, Estonia, Lithuania, Poland - from the EU sample - and Belarus). The reports of the European Commission (2018, 2019, 2020) contain a general summary of the condition of the EU insurance sector in a given year.

The presented research on the development and similarity of EU countries is innovative due to the methodology used - a multidimensional approach, and the long period of empirical research conducted.

**Conclusions**

Development paths and cluster analysis methods that we exploit and that have not been used in the literature, made it possible to synthetically summarize the multidimensional analysis of the development and similarity of sectors over the past 15 years. From the analysis performed we can draw the following conclusions, which confirm the hypothesis:

1. Countries that joined the Union in 2004. have not yet overcome the differences in the structure of insurance sectors. The development of each group of old and new EU countries is illustrated in Figures 1.1-1.14. Fig. 1.13 is a picture of development paths which show that each country did not change the level of development of the insurance sector after 2004.
2. In each of the years under consideration, the countries that joined the EU in 2004 belong to the same groups (the only exception here is Malta and Cyprus in 2005) (cf. Fig. 2.1). Moreover, these groups are characterised by the highest Silhouette clustering quality index (cf. Fig. 2.3).
   This indicates a similarity between the insurance markets of these countries in the period studied. We can say that after 15 years of membership in the UE the insurance markets of Poland, Estonia, Latvia, Slovakia, the Czech Republic, Hungary, Slovenia, Malta and Cyprus still form a separate group and remain very similar one to another. When compared to the markets of the old Union they are closest to Greece, Portugal and Spain.
3. In each of the years studied Luxemburg stays isolated. This stresses the distinctness of the insurance market among the markets of all the UE countries.

A high share of insurance in the market structure is generally treated as a certificate of maturity of this market. Insurance awareness is a very important element of a prosperous economy of the country. It allows citizens to use the available insurance offer, which protects against the risk of adverse fortuitous events. Demand for insurance largely depends on economic factors and the price of this service, i.e. the insurance premium.

The correct selection of factors is a condition for obtaining correct results in the study of the insurance structure of a given country. Between the groups of the old and the new European Union after 2004. there was a fairly large variation in the characteristics of insurance. In this work, multi-dimensional analysis was used to reveal similarities and differences between the individual countries in these groups of countries, in particular in 2005, 2009 and 2015, 2018. When examining the state of development of European Union insurance markets, it can be seen that still a long way separates Central and Eastern European countries from reaching western market levels. Nevertheless, the insurance industry is developing very dynamically in these countries. In each analyzed period, changes in the insurance structure of these countries can be seen. It is worth adding that the Polish insurance market has a great potential for further development.

**Acknowledgments**

This paper was financed by the Ministry of Science and Higher Education of the Republic of Poland as part of a research program "Regional Initiative of Excellence" Programme for 2019-2022. Project no.: 021/RID/2018/19. Total financing: 11 897 131.40 PLN

**Annex**

Fig. 1.1 Comparison of population (in thousands) of Old and New UE during 2004-2018

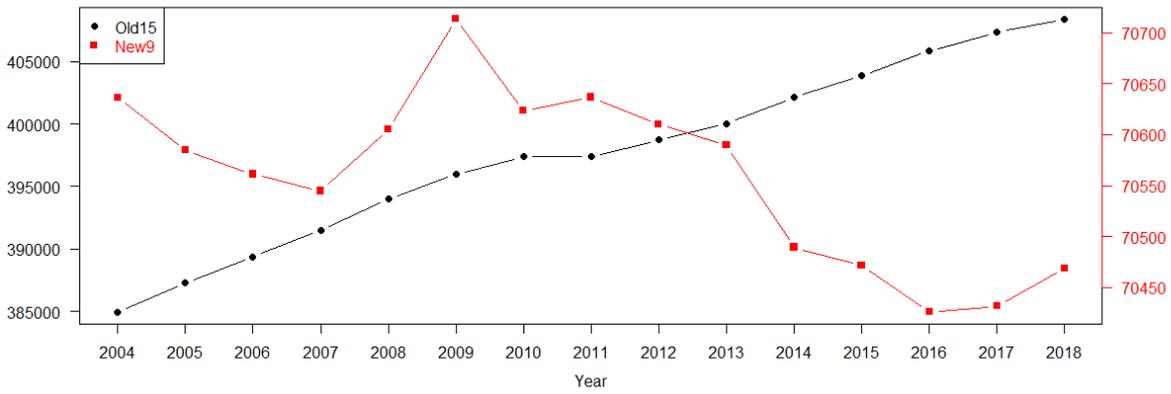

Source: Own analysis

Fig. 1.2 Number of companies for Old and New UE during 2004-2018

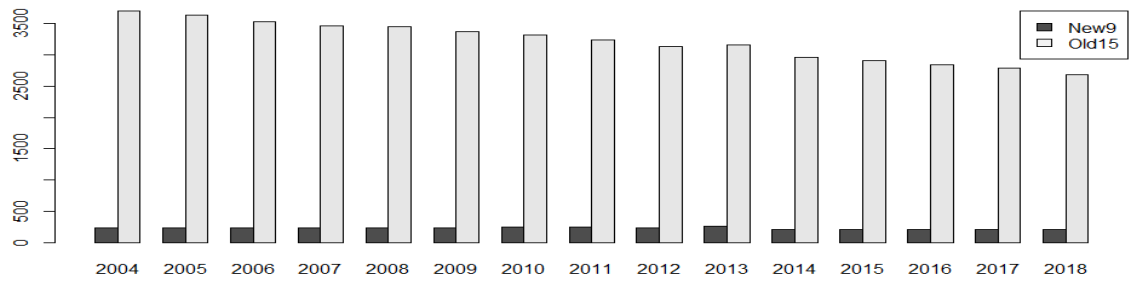

Source: Own analysis

Fig. 1.3 Assets Total in percent

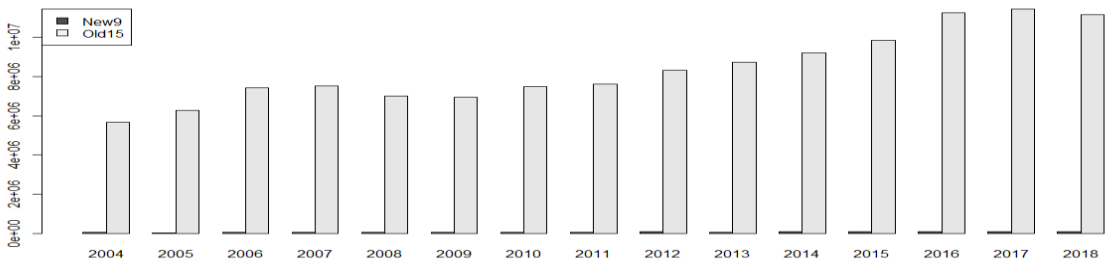

Source: Own analysis

Fig. 1.4 Total assets (EUR million) of Old and New UE during 2004-2018

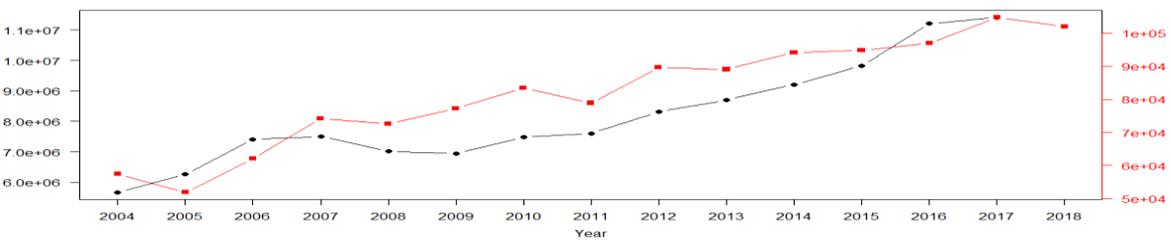

Source: Own analysis

Fig. 1.5 GDP (EUR million) of Old and New UE during 2004-2018

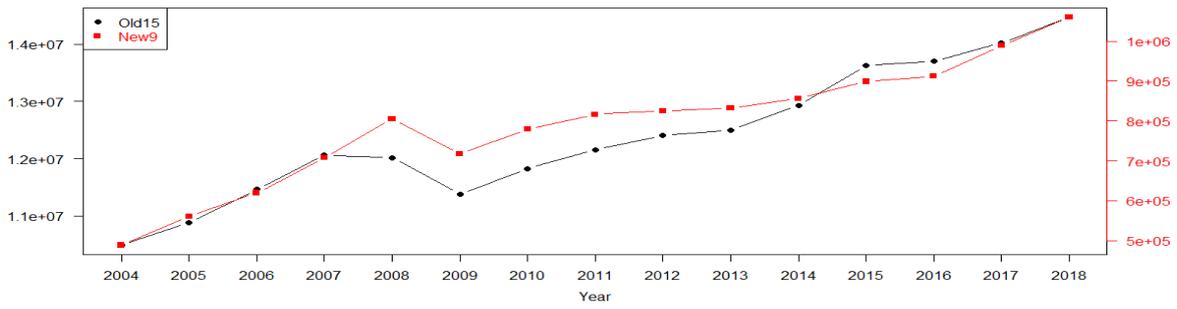

Source: Own analysis

Fig. 1.6 Gross Written Premium for Life, P&C, Health sector for Old and New EU during 2004-2018

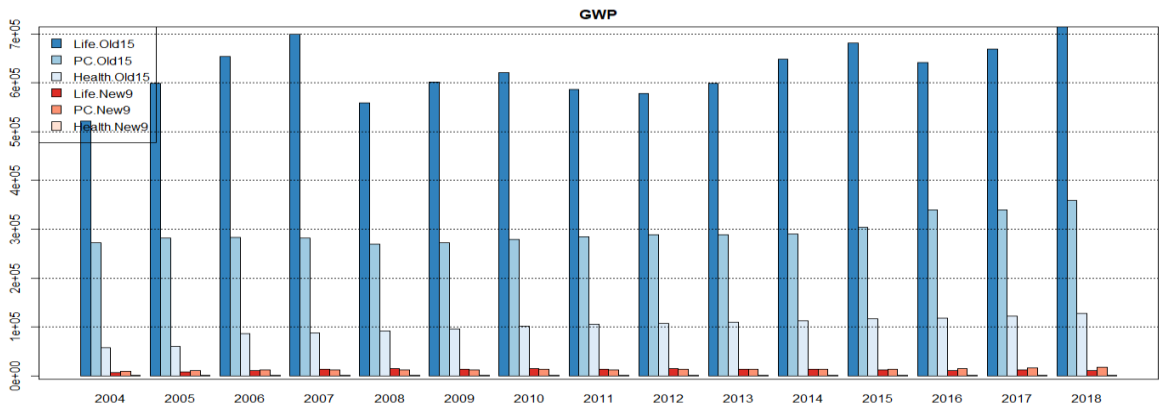

Source: Own analysis

Fig. 1.7 Gross Premium Written as a percentage for Life, P&C, Health sector for Old EU

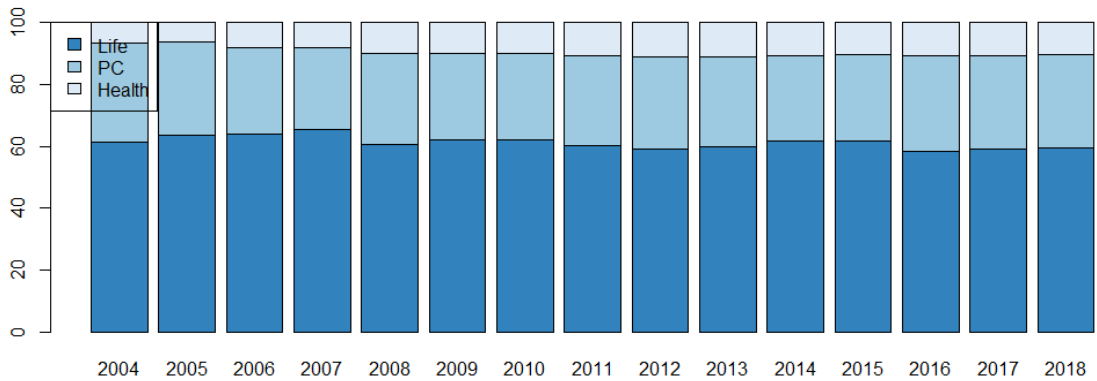

Source: Own analysis

Fig. 1.8 Gross Premium Written as a percentage for Life, P&C, Health sector for New EU

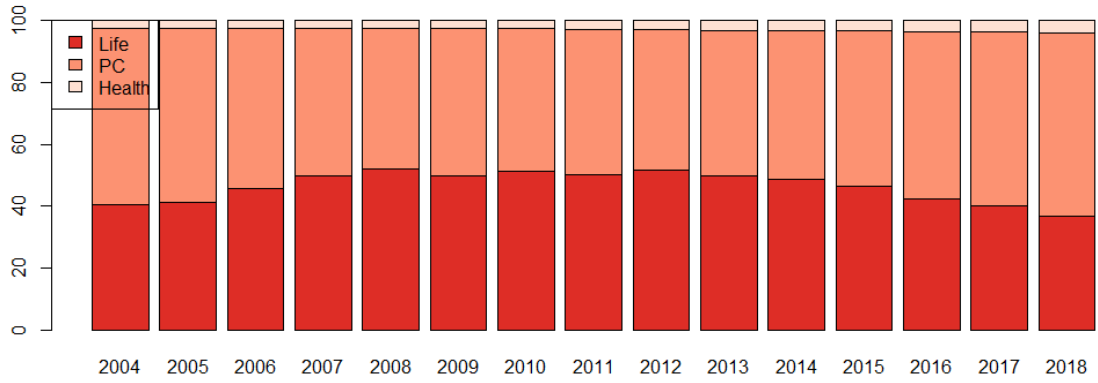

Source: Own analysis

Fig. 1.9 Penetration for Life, P&C and Health sector of Old and New EU during 2004-2018

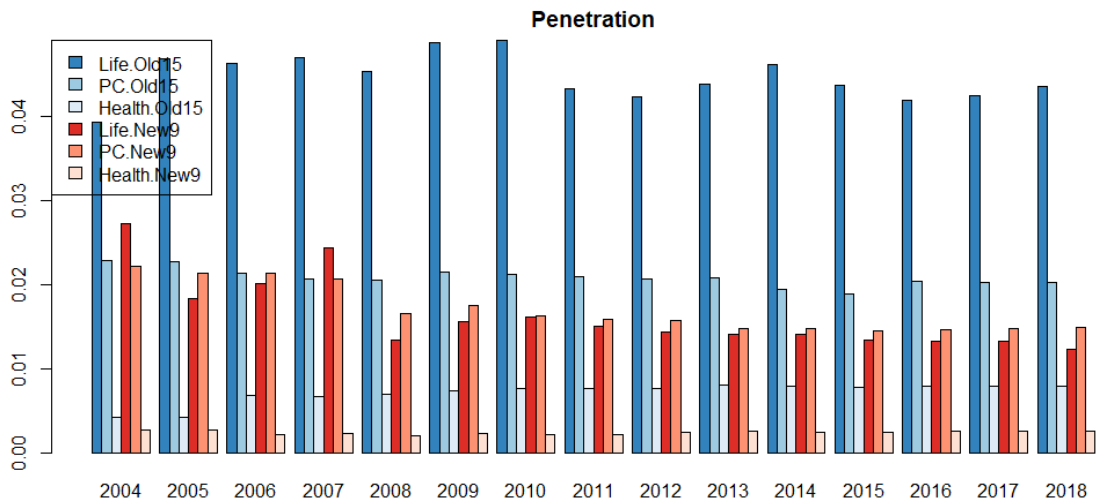

Source: Own analysis

Fig. 1.10 Density for Life, P&C and Health sector for Old and New EU

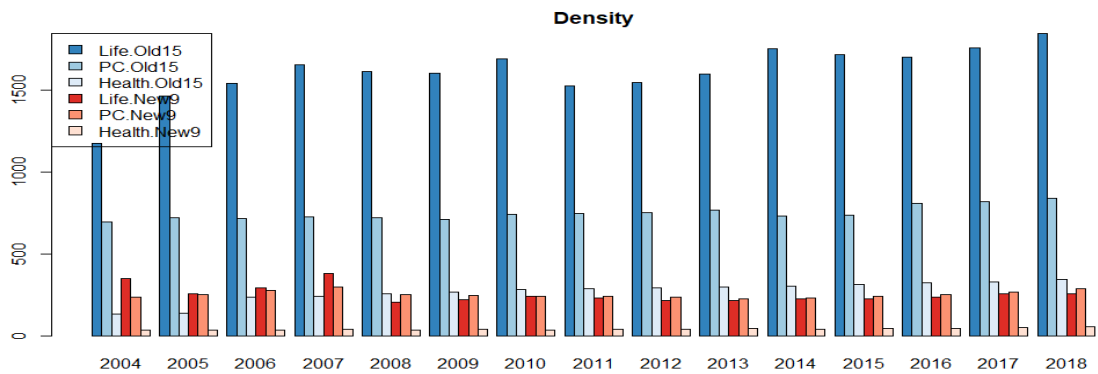

Source: Own analysis

Fig. 1.11 Investment portfolio on domestic market (EUR million) for Old and New EU

Source: Own analysis

Fig. 1.12 Market share of the top 5 for Life and P&C sector (country average) for Old and New EU

Source: Own analysis

Fig. 1.13 Development path of the Old and New EU during 2004-2018

Fig. 1.14 Ranking (X axis) and Hellwig's development measure (Y axis) for 2005, 2009, 2015 and 2018

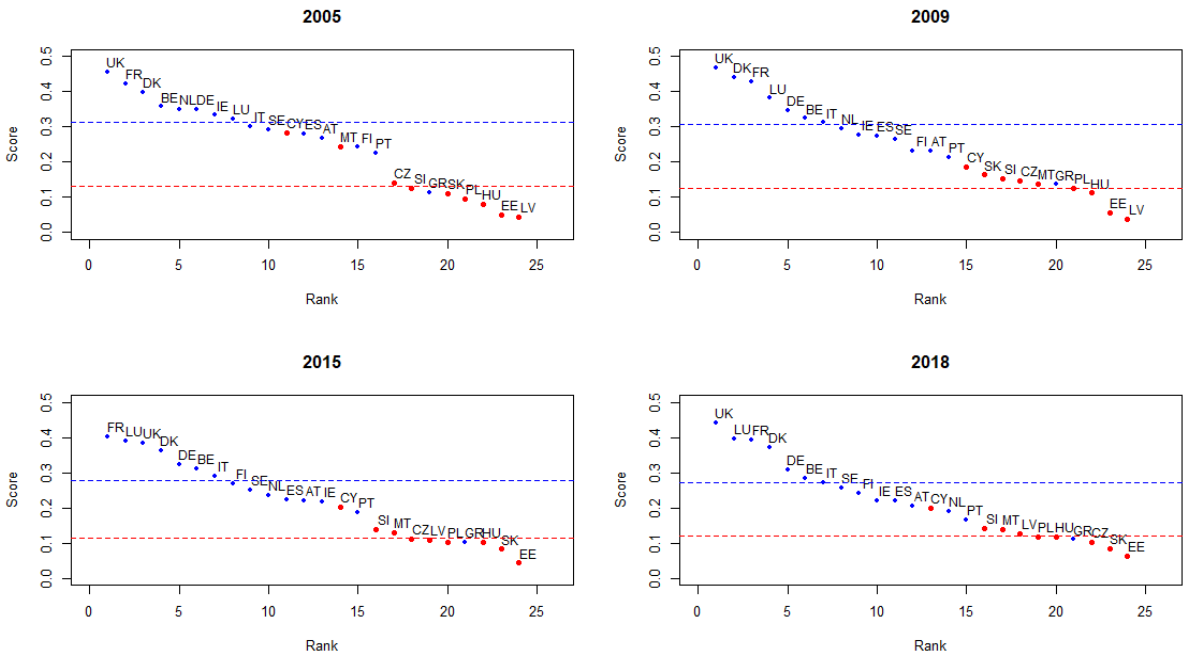

Source: Own calculations.

Fig. 1.15. Change in Hellwig's development measure in 2018 compared to 2004 (in %).

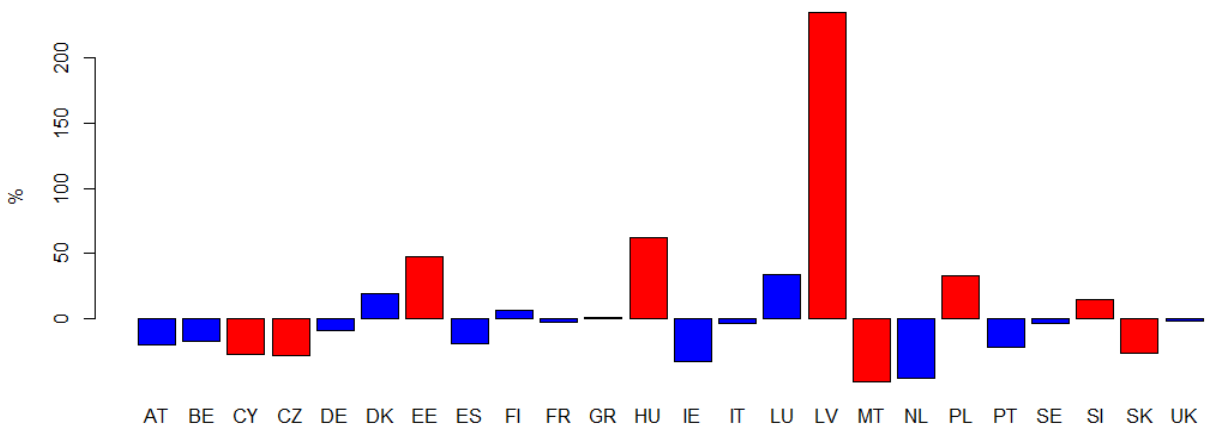

Source: Own calculations.

Table 2.1 Validation indices for data partitions for 2005, 2009, 2015 and 2018 year

| Year | 2005 | | | | | 2009 | | | | |
|---|---|---|---|---|---|---|---|---|---|---|
| | Number of clusters | | | | | Number of clusters | | | | |
| Validation criterion | 2 | 3 | 4 | 5 | 6 | 2 | 3 | 4 | 5 | 6 |
| | Ward's method | | | | | | | | | |
| Silhouette | 0,327 | **0,348** | 0,267 | 0,267 | 0,241 | 0,525 | **0,530** | 0,493 | 0,521 | 0,457 |
| Calinski Harabasz index | **12,929** | 12,204 | 11,540 | 10,994 | 10,684 | 8,241 | 6,871 | 8,749 | 9,426 | **9,749** |
| Dunn index | 0,326 | 0,460 | **0,465** | 0,445 | 0,425 | 0,230 | 0,230 | 0,308 | 0,390 | **0,449** |
| Xie-Beni index | 0,946 | 0,695 | 0,701 | **0,631** | 0,712 | 2,264 | 1,881 | 1,346 | 1,043 | **0,839** |
| | | | | | k-means | | | | | |
| Silhuette | **0,327** | 0,277 | 0,247 | 0,255 | 0,238 | 0,265 | **0,280** | 0,227 | 0,182 | 0,235 |
| Calinski Harabasz index | **12,929** | 10,465 | 9,134 | 11,177 | 10,883 | **11,022** | 10,421 | 10,003 | 7,228 | 9,191 |
| Dunn index | 0,326 | 0,275 | 0,267 | **0,465** | 0,425 | 0,195 | 0,402 | **0,451** | 0,297 | 0,378 |
| Xie-Beni index | 0,946 | 1,250 | 1,364 | **0,571** | 0,703 | 2,877 | 1,045 | **0,833** | 1,713 | 1,216 |
| | | | | | PAM | | | | | |
| Silhuette | 0,327 | **0,348** | 0,261 | 0,178 | 0,222 | 0,262 | **0,280** | 0,179 | 0,184 | 0,205 |
| Calinski Harabasz index | **12,929** | 12,204 | 11,182 | 9,332 | 10,072 | **10,765** | 10,421 | 7,837 | 8,529 | 8,353 |
| Dunn index | 0,326 | **0,460** | 0,426 | 0,336 | 0,393 | 0,281 | 0,402 | 0,385 | 0,432 | **0,441** |
| Xie-Beni index | 0,946 | 0,695 | **0,655** | 0,953 | 0,744 | 1,398 | 1,045 | 1,043 | 0,812 | **0,683** |
| Year | *2015* | | | | | *2018* | | | | |
| | Number of clusters | | | | | Number of clusters | | | | |
| Validation criterion | 2 | 3 | 4 | 5 | 6 | 2 | 3 | 4 | 5 | 6 |
| | | | | | Ward's method | | | | | |
| Silhouette | 0,265 | **0,274** | 0,229 | 0,160 | 0,131 | **0,656** | 0,537 | 0,496 | 0,595 | 0,575 |
| Calinski Harabasz index | **11,022** | 10,348 | 9,502 | 8,362 | 7,891 | 9,665 | **10,160** | 9,194 | 8,721 | 8,496 |
| Dunn index | 0,195 | 0,297 | **0,300** | 0,170 | 0,170 | 0,254 | **0,328** | 0,315 | 0,315 | 0,319 |
| Xie-Beni index | 2,877 | 2,175 | **1,781** | 5,304 | 4,587 | 1,436 | 1,050 | 1,376 | 1,154 | **1,029** |
| | | | | | k-means | | | | | |
| Silhuette | 0,273 | **0,276** | 0,122 | 0,186 | 0,210 | 0,300 | **0,310** | 0,189 | 0,176 | 0,221 |
| Calinski Harabasz index | 10,623 | 8,850 | 5,620 | 9,868 | **10,745** | 10,805 | **11,003** | 9,806 | 9,609 | 10,716 |
| Dunn index | 0,215 | 0,225 | 0,115 | 0,187 | **0,245** | 0,192 | **0,289** | 0,162 | 0,244 | 0,206 |
| Xie-Beni index | 2,152 | **1,896** | 6,082 | 3,642 | 2,813 | 2,410 | **1,755** | 4,612 | 2,262 | 2,865 |
| | | | | | PAM | | | | | |
| Silhuette | 0,253 | **0,263** | 0,245 | 0,157 | 0,121 | 0,208 | **0,216** | 0,179 | 0,134 | 0,137 |
| Calinski Harabasz index | 9,842 | **10,821** | 9,855 | 8,294 | 7,484 | 8,604 | **9,215** | 8,957 | 7,970 | 7,374 |
| Dunn index | 0,221 | 0,328 | **0,344** | 0,330 | 0,344 | 0,201 | 0,263 | **0,303** | 0,295 | 0,289 |
| Xie-Beni index | 2,007 | 1,431 | 1,172 | 1,263 | **1,036** | 2,353 | 1,743 | 1,397 | 1,291 | **1,179** |

Source: Own calculations performed with the use of the 'clusterSim' package developed by M. Walesiak and A. Dudek (Silhouette and Calinski Harabasz index) and the 'clusterCrit' package developed by Bernard Desgraupes (Dunn and Xie-Beni index).
Note: numbers in bold indicate the optimal number of groups with reference to a given criterion

Fig. 2.1. Dendrogram for Ward's method clustering for 2005, 2009, 2015 and 2018

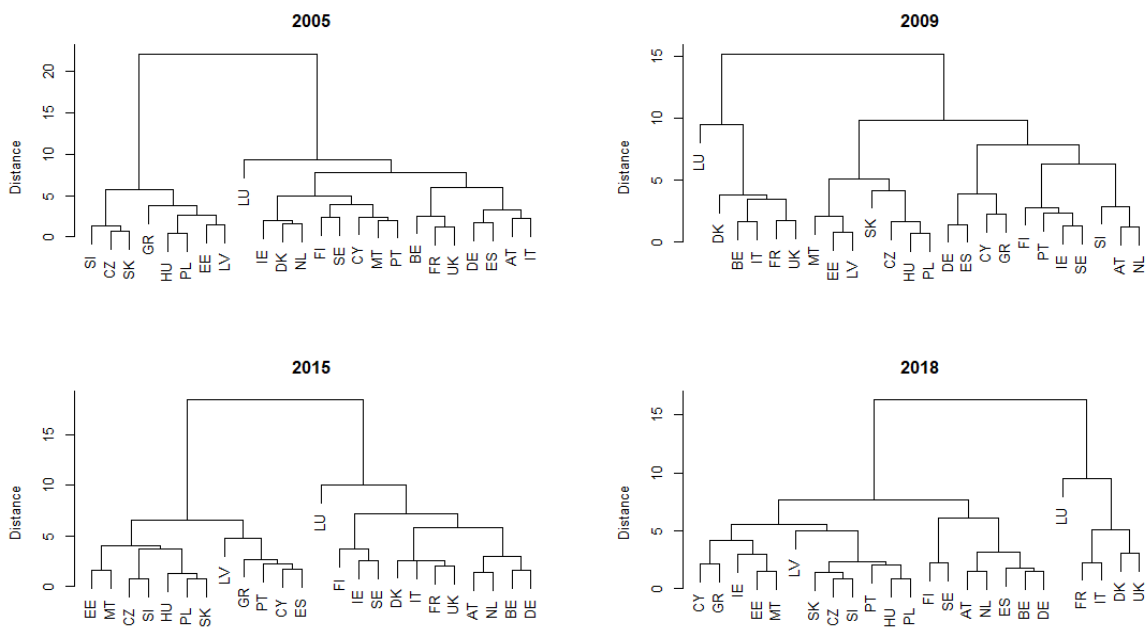

Source: Own calculations.

Fig. 2.2 Cluster analysis results for 2005, 2009, 2015 and 2018

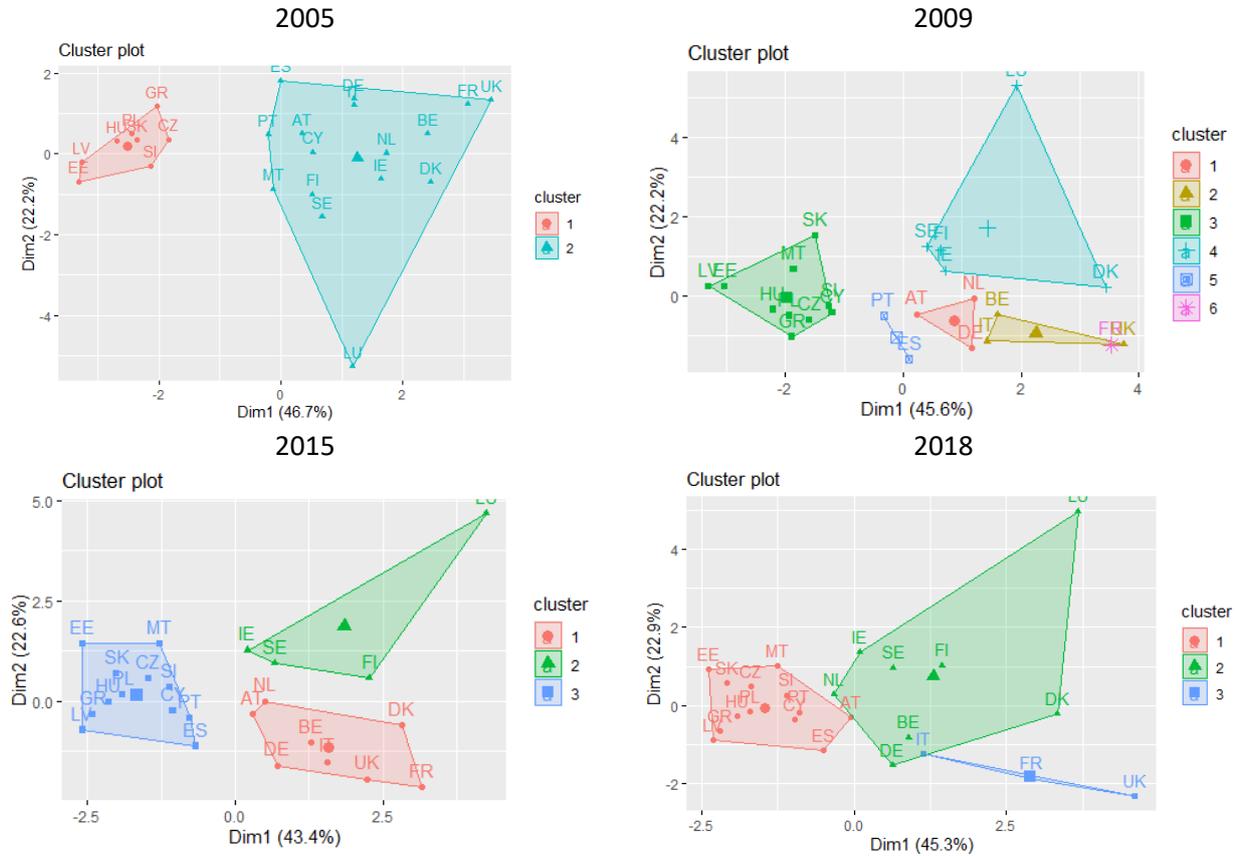

Source: Own calculations.

Fig. 2.3 Silhouette plot for clustering in 2005, 2009, 2015 and 2018

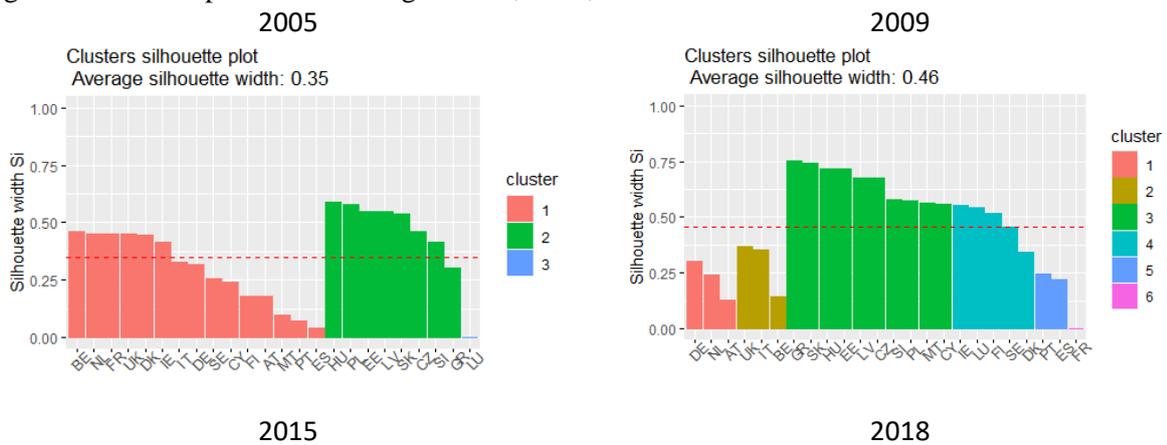

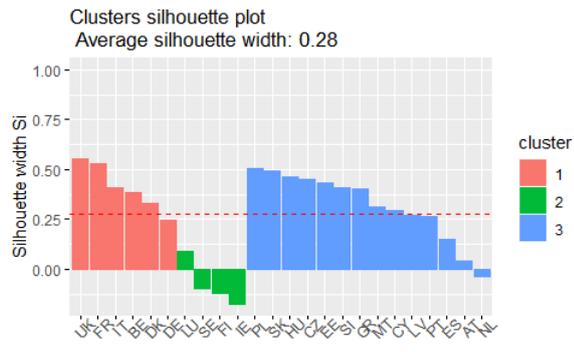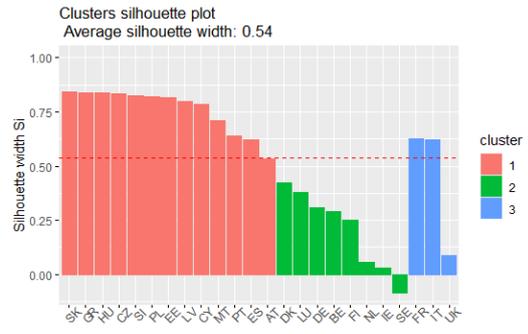

Source: Own calculations.